\documentclass[a4,aps,twocolumn,showpacs,preprintnumbers,amslatex,amsmath,amssymb]{revtex4-1}

\usepackage{fancyhdr}
\usepackage{epsfig}
\usepackage{graphicx}
\usepackage{verbatim}
\usepackage{indentfirst}
\usepackage{latexsym}
\usepackage{dsfont}

\usepackage{color,soul}

\usepackage{titlesec}
\usepackage{colortbl}

\usepackage[colorlinks,breaklinks]{hyperref}
\usepackage{hyphenat}
\usepackage{natbib}

\begin{document}

\title{Symmetries and entanglement in the one-dimensional spin-1/2 XXZ model}
\author{Mykhailo V. Rakov$^1$, Michael Weyrauch$^2$, Briiissuurs Braiorr-Orrs$^2$}
\affiliation{$^1$ Kyiv National Taras Shevchenko University, 64/13 Volodymyrska st., Kyiv 01601, Ukraine}
\affiliation{$^2$ Physikalisch-Technische Bundesanstalt, Bundesallee 100, D-38116 Braunschweig, Germany}

\begin{abstract}
  An efficient and stable algorithm for U(1) symmetric matrix product states (MPS) with periodic boundary conditions (PBC) is proposed. It is applied to a study of correlation and entanglement properties of the eigenstates of the spin-1/2 XXZ model with different spin projections.  Convergence properties and accuracy of the algorithm are studied in detail.
\end{abstract}


\maketitle

\section{Introduction}

Tensor networks and, more specifically, matrix product states (MPS) are convenient ways to represent quantum states.
By now, the available literature on this subject is vast, and many different algorithms based on tensor network representations have been proposed and implemented. In particular, the extremely successful DMRG algorithm~\cite{WHI93b} has been rephrased in MPS language~\cite{DUK98}, and modern implementations of DMRG use MPS representations. For a detailed review see e.g. Ref.~\cite{SCH10}.

The algorithms reviewed in Ref.~\cite{SCH10} use non-symmetric MPS. However, due to the Mermin-Wagner theorem a continuous symmetry cannot be broken~\cite{PhysRevLett.17.1133} in one dimension (1D).  Therefore, for physical as well as numerical reasons it is desirable to construct MPS respecting symmetries, e.g. U(1) or SU(2) symmetry, as dictated by the physical problem under consideration. In fact, SU(2) symmetric
MPS have been used already in the early MPS papers by {\"O}stlund and Rommer~\cite{OEST95, OST1997} in order to optimize the number of MPS parameters to be determined. McCulloch discussed practical issues related to the MPS implementation for Abelian and non-Abelian symmetries~\cite{MCC2007}. More recently, Vidal and collaborators provided a rather systematic presentation of symmetries in tensor networks. In a series of papers~\cite{singh:1,PhysRevB.83.115125, SIN2012} the essential structure  of symmetric tensor network states was clarified.

In the present paper, we propose an efficient and stable algorithm
for U(1) symmetric MPS for periodic boundary conditions (PBC). More specifically, we modify the PBC algorithm suggested by Verstraete, Porras, and Cirac~\cite{VER04} and augment it by a novel method to construct U(1) symmetric MPS. Ground or excited states with any desired spin projection can be targeted easily. Theoretical and practical aspects not covered in the more general papers cited above will be addressed and the differences to the more standard open boundary condition (OBC) implementations will be stressed.

We test this algorithm with a rather detailed study of the spin-1/2 XXZ model in an external magnetic field $h$
\begin{equation}\label{XXZmodeleq}
H=\sum_{i=1}^N (s_i^x \otimes s_{i+1}^x + s_i^y \otimes s_{i+1}^y + \Delta \, s_i^z \otimes s_{i+1}^z)+h \,\sum_{i=1}^N s_i^z,
\end{equation}
where the index $N+1$ is set to $1$. The spin operators $s_i^{\alpha}$ ($\alpha=x,y,z$) are related to the Pauli matrices $\sigma_i^\alpha$ by $s_i^{\alpha}=\sigma_i^{\alpha}/2$. The parameters of the model are the anisotropy parameter $\Delta$ and the magnetic field $h$.

This model is U(1) symmetric, i.e. its Hamiltonian $H$ commutes with the $z$-component $ S_z=\sum_{i=1}^N  s_i^z$ of the total spin operator ${\vec S}$. Furthermore, it is $Z_2$ spin-reflection symmetric. The XXZ model can be solved using the Bethe Ansatz~\cite{PhysRev.147.303,PhysRev.150.321,PhysRev.150.327,PhysRev.151.258,PhysRevA.80.032102, KAR98}. These Bethe Ansatz results will serve as a convenient benchmark.

\begin{figure}[h!]
    \includegraphics[width=0.45\textwidth]{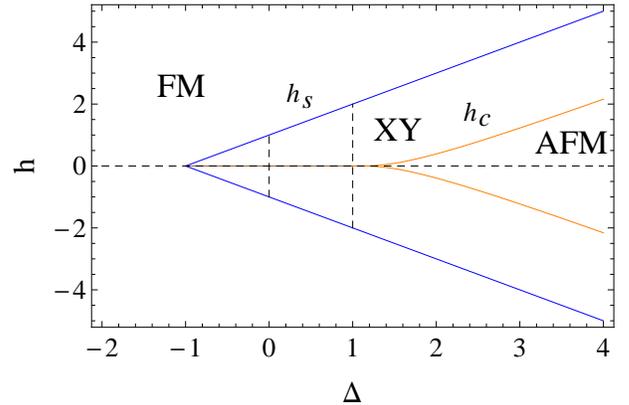}
    \caption{\footnotesize (color online)
    Phase diagram of the spin-1/2 XXZ model in a magnetic field along $z$ axis.
    The critical line $h_s$ separates the spin liquid (XY) phase from the ferromagnetic (FM) phase. The critical line $h_c$ separates the XY phase from the anti-ferromagnetic (AFM) phase. We investigate the model in detail along the dashed lines.
    } \label{phased}
\end{figure}

The ground state phase diagram of the spin-1/2 XXZ model as obtained from the Bethe Ansatz is shown in Fig.~\ref{phased}.
There are three phases: ferromagnetic (FM), spin-liquid (XY),
and anti-ferromagnetic (AFM). These three phases are separated by two lines, $h_s=1+\Delta$ and $h_c$ (see Ref. \cite{PhysRev.151.258}, Eq.(8)). At $h=0$ the XXZ system undergoes a first-order phase transition at $\Delta=-1$ and a Kosterlitz-Thouless infinite-order phase transition at $\Delta=1$~\cite{PhysRevA.85.052128}. The line between
these two points is a critical line, where the excitation gap vanishes. This line separates different spin liquid phases. Specifically, we study the model along the lines indicated by dashes in the phase diagram.

The numerical MPS solution provides an explicit representation of the wave functions.
Due to the U(1) symmetry the wave functions are simultaneously eigenstates of the Hamiltonian and of $S_z$.
As a consequence, the magnetization $m_z={\langle S_z \rangle}/{N}$ can be used as a quantum number to label the states.
We determine properties of the XXZ system with different magnetizations $m_z$ as a function of the anisotropy parameter $\Delta$. {(Alternatively, they could be determined as functions of $\Delta$ and the magnetic field $h$.)
 } It turns out that these states have interesting entanglement properties. In fact, the amount, range, and type of entanglement determines if a state can be successfully modeled by an MPS of a given size. In order to study this quantitatively we will calculate various entanglement quantifiers.

 We compare our calculations for 50 and 100 spins with finite-size Bethe Ansatz results. Since spin systems of 100 sites are relatively close to the thermodynamic limit, we also compare with analytical infinite-size Bethe Ansatz results. Furthermore, we discuss
convergence problems in detail: convergence to the desired state depends not only on the matrix size $m$
of the MPS but also on the choice of the U(1) symmetry sectors and their degeneracies. Moreover, as our implementation uses the facility introduced in Ref.~\cite{PIP10} to represent `long' products of large transfer matrices by eventually rather small singular value decompositions, we will study in some detail how this facility can be used profitably in practice. Experience shows that one has to be extremely careful in order not to choose the size of the singular value decomposition too small. In fact, we do not share the positive experience made in Ref.~\cite{PIP10} for spin-1 Heisenberg systems.

The paper is organized as follows. In section~\ref{standard} we outline the PBC MPS formalism used in this paper. Our novel implementation of U(1) symmetric MPS is presented in section~\ref{U1MPS}.
Application of this algorithm to the 1D spin-1/2 XXZ model together with comparisons to Bethe Ansatz
calculations is presented in section~\ref{XXZmodel}. Finite size and convergence issues are also discussed there. A few infinite-size Bethe Ansatz results are listed in the Appendix A.

\section{MPS formalism for PBC}\label{standard}

Here we review the PBC formalism proposed by Verstraete, Porras, and Cirac (VPC)~\cite{VER04}.  We include a number of modifications such as the use of matrix product operators (MPO) and a circular and efficient local update as first suggested by Pippan, White and Evertz (PWE)~\cite{PIP10}. In this and in the next section we denote $s_i^z$ simply as $s_i$ to avoid a large number of indices.

The state of a 1D quantum spin system of size $N$ is approximated in terms of a matrix product state
\begin{equation}\label{eq:MPS}
|\psi\rangle=\sum_{\vec s} {\rm Tr}\; M^{[1],s_1}\cdot
\ldots \cdot M^{[N],s_N} |s_1 \dots s_N\rangle.
\end{equation}
Here, the $s_i$ represent the local degrees of freedom at the site
$i$, and each $M^{[i],s_i}$ represents a matrix of size $m
\times m$, where $m$ is called bond dimension, i.e., $M^{[i]}$ is a rank-3 tensor. In the algorithm to
be described the elements of these tensors  $M_{a_{i-1},a_i}^{[i],s_i}$ (with $a_0=a_N$) are variational
parameters to be adjusted using a suitable optimization procedure.

Analogously, any operator is written as a matrix product operator
\begin{equation}\label{eq:MPO}
O=\sum_{\vec s,\vec s^{\prime}} {\rm Tr}\; W^{[1],s_1,s_1^\prime} \ldots W^{[N],s_N,s_N^\prime}
|s_1 \dots s_N \rangle\langle s_1^\prime \dots s_N^\prime|.
\end{equation}
Again, each $W^{[i],s_i,s_i^\prime}$
represents a matrix of size $m_W \times m_W$, i.e. each $W^{[i]}$ is a rank-4
tensor with elements $W_{b_{i-1},b_i}^{[i],s_i,s_i^{\prime}}$ (with $b_0=b_N$). In particular, the MPO representation of the XXZ Hamiltonian  given in Eq.~(\ref{XXZmodeleq}) consists of the following rank-4 tensors,
\begin{eqnarray}
\hat W^{[1]}&=&\begin{pmatrix} h s_1^z & s_1^x & s_1^y & \Delta s_1^z & \mathds{1} \\ 0 & 0 & 0 & 0 & s_1^x \\ 0 & 0 & 0 & 0 & s_1^y \\ 0 & 0 & 0 & 0 & s_1^z \\ 0 & 0 & 0 & 0 & 0 \end{pmatrix}, \\
\hat W^{[i]}&=&\begin{pmatrix} \mathds{1} & 0 & 0 & 0 & 0 \\ s_i^x & 0 & 0 & 0 & 0 \\ s_i^y & 0 & 0 & 0 & 0 \\ s_i^z & 0 & 0 & 0 & 0 \\ h s_i^z & s_i^x & s_i^y & \Delta s_i^z & \mathds{1} \end{pmatrix}, \hspace{0.05 cm} i=2,\dots,N.
\end{eqnarray}

Matrix elements of an MPO in MPS
\begin{equation}\label{eq:matrixe}
\langle\phi|O|\psi\rangle={\rm Tr}\; E_W^{[1]}(A,B)\cdot\ldots \cdot E_W^{[N]}(A,B)
\end{equation}
can be conveniently expressed in terms of the (generalized) transfer matrices
\begin{equation}\label{eq-transfer1}
E_W^{[i]}(A,B) = \sum_{s_i,s_i^\prime} \, W^{[i],s_i,s_i^\prime} \otimes (B^{[i],s_i})^*\otimes
 A^{[i],s_i^\prime}.
\end{equation}
The tensors $B$ and $A$ characterize the states $|\phi\rangle$ and
$|\psi\rangle$, respectively.  The Kronecker product $\otimes$ in
Eq.~(\ref{eq-transfer1}) obviously produces matrices of size
$ m_W m^2 \times m_W m^2$.
The special transfer matrix $E_1^{[i]}(A,B)$ represents the matrix element $\langle\phi|\psi\rangle$ of the identity operator.

In order to find the ground state of a many body system one solves
a standard variational problem using the matrix elements of the MPS
as variational parameters. The optimization of the variational
parameters of the MPS is implemented as a local update step, which
is repeated until convergence is achieved~\cite{VER04}.
In the MPO formalism for PBC such a local update step amounts to the solution of a
generalized eigenvalue problem
\begin{equation}\label{eq:eigenproblem-p}
H_{\rm eff}^{[i]} \, \nu^{[i]}=\epsilon^{[i]} \, N_{\rm eff}^{[i]} \, \nu^{[i]}
\end{equation}
in terms of the effective Hamiltonian $H_{\rm eff}^{[i]}$ and the
effective normalization matrix $N_{\rm eff}^{[i]}$ given by
\begin{eqnarray}
H_{\rm eff}^{[i]}&=&\sum_{k,l=1}^{m_W} W^{[i]}_{kl} \otimes \widetilde{\left(({H_R^{[i]} \cdot H_L^{[i]}})_{lk}\right)},\label{eq:eff-Hamiltonian-p}\\
N_{\rm eff}^{[i]}&=&\mathds{1} \otimes \left(\widetilde{N_R^{[i]}\cdot N_L^{[i]}}\right)\label{eq:eff-norm-p}.
\end{eqnarray}
The matrices $H_L^{[i]},~N_L^{[i]}$ and  $H_R^{[i]},~N_R^{[i]} $ are
the products of transfer matrices from all sites to the left and to
the right of the site $i$, where the MPS is updated. (In order to define which sites are left or right of a site $i$ one initially arbitrarily numbers all sites from 1 to $N$, and sites with $j<i$ are left and sites with $j>i$ are right of site $i$.) The $H$ and $N$
matrices are obtained from transfer operators $E^{[j]}$ as defined
in Eq.~(\ref{eq-transfer1}) with the MPO of the Hamiltonian for $H$ and the unity MPO for $N$, in both
cases setting $A=B=M^{[j]}$.

The tilde
in~(\ref{eq:eff-Hamiltonian-p}) and~(\ref{eq:eff-norm-p}) indicates the operation
$X_{(ij),(kl)}=\tilde{X}_{(ik),(jl)}$
for each $m^2 \times m^2$ matrix. As
a consequence of this transposition the effective Hamiltonian and
the normalization matrix are assured to be Hermitian matrices and
standard methods for the solution of generalized eigenvalue problems
can be applied.

The energy of the state is obtained from $\epsilon^{[i]}$, and this
value will converge to the ground state energy eventually. In fact,
we stop the iterative update procedure, if this quantity does not
change any more with respect to defined convergence criteria.
The updated MPS is obtained from the generalized eigenvector
\begin{equation}\label{newmps}
\nu_{(s_i,a_i,a_{i-1})}^{[i]}=M_{a_{i-1},a_i}^{[i],s_i}
\end{equation}
by a suitable partitioning
of the vector into a tensor.

The tensors $H_L^{[i]},~N_L^{[i]}$, $H_R^{[i]},~N_R^{[i]}$ can be calculated in different ways. In the VPC approach~\cite{VER04,SCH10} one sweeps back and forth over the entire system. The tensors are calculated straightforwardly by successive multiplication by the appropriate transfer matrix $E_W^{[i]}$ or $E_1^{[i]}$, starting from the leftmost and rightmost sites of the system, respectively. In the PWE approach~\cite{PIP10,weyrauchrakov,1742-5468-2011-05-P05021} one subdivides the system into three sections and optimizes the MPS always from left to right in each section and `moves' (updates) in a circle.

The PWE approach is able to take advantage of the fact that `long' products of transfer matrices  have singular values that may decay rather fast.
In the PWE approach the minimum length of a product of transfer matrices is $N/3$, so that for a system with size of about 100 spins this length may already be `long'.  Thus $H_L^{[i]},~N_L^{[i]}$, $H_R^{[i]},~N_R^{[i]}$ may be replaced by their singular value decomposition (SVD) with only a small number of singular values kept, thus dramatically reducing the computational resources required to calculate these tensors. The number of singular values we keep is called $p$ for $N$-tensors and $p^{\prime}$ for $H$-tensors. One finds that $p$ and $p^{\prime}$ depend approximately linearly on the bond dimension $m$~\cite{weyrauchrakov,1742-5468-2011-05-P05021}. Our experience shows that the PWE method has to be used with caution in order to prevent the algorithm from becoming unstable. We will comment on this further in section~\ref{XXZmodel}.

Whatever update strategy is used, one runs over the entire system several times updating the MPS at each site until convergence of the energy $\epsilon^{[i]}$ is achieved. Initially,
one starts from a randomly selected MPS. After each update step the
local MPS tensor is regauged in order to keep the algorithm stable.
This means, we have to assure that one of the following relations hold for each local tensor
\begin{eqnarray}\label{LRnormalization}
Q^L&=&\sum_{s_i}  M^{[i],s_i \dag} M^{[i],s_i}=\mathds{1}, \text{left-normalization} \nonumber\\
Q^R&=&\sum_{s_i}  M^{[i],s_i} M^{[i],s_i \dag}=\mathds{1}, \text{right-normalization}
\end{eqnarray}
This is possible because local MPS tensors are only defined up to a gauge freedom.

We would like to mention that it is easily possible to construct excited states along similar lines
 by finding
the lowest state in the space orthogonal to the space spanned by the
states already found~\cite{POR06,1367-2630-14-12-125015}.

\section{U(1) covariant MPS}\label{U1MPS}

In this section we
construct U(1) symmetric MPS. First we describe the construction of U(1) invariant MPS (with spin projection $S_z=0$) and then
covariant MPS with given $S_z$.
The approach is general and applies to any U(1) symmetric system (i.e., not only to a spin system).

The construction of MPS invariant under a symmetry is described in detail in many papers (see, e.g.,~\cite{PhysRevB.83.115125}). Each local tensor decomposes into a structural part and a degeneracy part according to the Wigner-Eckart theorem. Thus for U(1) symmetry the bond indices decompose into a spin projection index and a degeneracy index: $a_i=\{m_i,\alpha_i\}$; where $\alpha_i=1$ through $t_{m_i}$ enumerate the degeneracy of a particular $m_i$. In practice one has to choose appropriate finite sets $\{m_i\}$ with corresponding $\{t_{m_i}\}$. They are not determined by symmetry;
this fact introduces significant additional freedom into the algorithm.

For U(1) symmetry the Wigner-Eckart theorem takes a very simple form, and the matrix elements are given by
\begin{equation}\label{mel}
M_{(m_{i-1},\alpha_{i-1}),(m_i,\alpha_i)}^{[i],s_i}=T_{(m_{i-1},\alpha_{i-1}),(m_i,\alpha_i)}^{[i]} \cdot \delta_{m_{i-1},m_i+s_i}.
\end{equation}
The matrix elements $T_{(m_{i-1},\alpha_{i-1}),(m_i,\alpha_i)}^{[i]}$ of the degeneracy part are often called  `reduced matrix elements'. In the case of U(1) symmetry the reduced matrix elements are equal to the standard matrix elements, if the latter are nonzero.

Alternatively, it may be said that the local tensors decompose into a block structure, and the positions of the nonzero blocks are determined by the `conservation law'
\begin{equation}
m_{i-1}=m_i+s_i,
\end{equation}
while the size of the blocks is determined by the degeneracy indices.
%

The construction~(\ref{mel}) of U(1) symmetric matrices encodes the symmetry information within the matrix layout. No separate `quantum number' labels are required. If we want to maintain this property for the construction of the algorithm then
for PBC the leftmost and the rightmost indices must be the same, and the procedure described above only constructs U(1) invariant states, i.e. states with $S_z=0$.
The clue for the practical construction of U(1) covariant MPS for PBC is obtained from Refs.~\cite{MCC2001,MCC2007,PhysRevB.83.115125}: a fictitious charge $-S_z$ (i.e., a non-interacting spin with spin projection $-S_z$) is inserted into the system at an arbitrary position. The modified system has total spin projection $S_z=0$ and can be described by a U(1) invariant tensor network.

For convenience, let us insert the fictitious charge at site $N+1$, i.e. between site $N$ and site 1. The tensor at the new site is a single matrix (because $s_{N+1}=-S_z$), and its matrix elements are
\begin{multline}
M_{(m_N,\alpha_N),(m_{N+1},\alpha_{N+1})}^{[N+1],-S_z}=\\
=T_{(m_N,\alpha_N),(m_{N+1},\alpha_{N+1})}^{[N+1]} \cdot \delta_{m_N,m_{N+1}-S_z}.
\end{multline}
The corresponding MPO at this fictitious site is just a unity MPO, since the site should be non-interacting.
Using this modified MPS one determines a U(1) invariant state and its corresponding energy as described in the previous section.

In order to find the required U(1) covariant state one eliminates the `fictitious charge' by multiplying its matrix into the tensor of a neighboring site, e.g. each of the matrices $M^{[N],s_N}$ is multiplied to matrix $M^{[N+1],-S_z}$.

Then,
the matrix elements of the new tensor $M^{\prime \, [N]}$
\begin{multline}
\hat M_{(m_{N-1},\alpha_{N-1}),(m_N,\alpha_N)}^{\prime \, [N], s_N}=\\
=T_{(m_{N-1},\alpha_{N-1}),(m_N,\alpha_N)}^{\prime \, [N]} \cdot \delta_{m_{N-1},m_N+s_N-S_z}
\end{multline}
fulfill the `conservation law'
\begin{equation}\label{conslawnew}
m_{N-1}+S_z=m_N+s_N.
\end{equation}
It can be easily checked explicitly that the resulting MPS has spin projection $S_z$ as required.
The conservation law (\ref{conslawnew}) is different from the conservation law (\ref{mel}) fulfilled at the other sites of the system.

The matrix $\hat M^{[N+1],-S_z}$ is strongly off-diagonal for large $|S_z|$. As a consequence, for a given $m$ and $S_z$ this matrix may vanish, i.e. cannot be constructed.  E.g., to construct a random MPS for $m_z=1/2$ for a system of $N$ sites one needs at least $N+1$ degeneracy sectors, which is prohibitive for practical calculations. For smaller $m_z$ the minimal number of required degeneracy sectors is smaller, but unlike the $m_z=1/2$ state these states are strongly entangled and need enough parameters for a suitable representation. As a consequence, the algorithm may converge to a wrong energy or get unstable: the MPS would be a bad variational Ansatz with too few parameters.

Here, we propose a way for the construction of U(1) covariant MPS for PBC that does not run into such problems. In fact, we propose to insert fictitious charges at several sites within the system.
This leads to an MPS with the following matrix elements,
\begin{equation}\label{me}
M_{(m_{i-1},\alpha_{i-1}),(m_i,\alpha_i)}^{[i],s_i}=T_{(m_{i-1},\alpha_{i-1}),(m_i,\alpha_i)}^{[i]} \delta_{m_{i-1}+x_i,m_i+s_i}
\end{equation}
with $x_i$ fixed at each site and $\sum_{i=1}^N x_i=S_z$. The Kronecker delta in Eq.~(\ref{me}) implies that the $x_i$ can only be half-integer or integer.

It can be easily checked by insertion into Eq.~(\ref{eq:MPS}) that the matrices defined in Eq.~(\ref{me}) produce an MPS with the desired spin projection $S_z$.
The difference between this approach and the (naive) approach described above is that the total charge $S_z$ is distributed among all spins. This can be done because U(1) symmetry is Abelian.

The `conservation law' to be fulfilled at each site
\begin{equation}\label{conslaw}
m_{i-1}+x_i=m_i+s_i
\end{equation}
must be supplemented with the condition $\sum_i x_i=S_z$. For $m_z=1/2$ the choice of $x_i$ is obvious: $x_i=s_i=1/2$ at each site, and only one degeneracy sector for the virtual indices is needed. But for $m_z<1/2$ these conditions can be fulfilled in various ways. We choose one of them, which distributes $S_z$ over all spins as homogeneously as possible. To this end, $S_z$ is split into small portions, namely 1/2 for spin-1/2 systems.

Since $|S_z| \le \frac{N}{2}$ for spin-1/2 systems, $x_i=0$ for a certain number of sites and $x_i=\frac{1}{2}$ for the others. We place all nonzero $x_i$ at one end of the system and all zero $x_i$ at the other end (with respect to our enumeration $1,\ldots,N$ of the sites).
Thus, the conservation laws are
\begin{eqnarray}
m_{i-1}&=&m_i+s_i-{\rm Sgn}(S_z) \cdot \frac{1}{2},~~~{\rm for}~  i \le 2|S_z| \\
m_{i-1}&=&m_i+s_i,~~~~{\rm for}~   i > 2|S_z|.
\end{eqnarray}

The MPS matrices are block main/lower/upper diagonal. Let us introduce the following notations:  $\textsf{d} \equiv$ block diagonal, $\textsf{ld}, \textsf{lld}, \dots \equiv$ block 1st, 2nd, \dots lower diagonal, $\textsf{ud}, \textsf{uud}, \dots \equiv$ block 1st, 2nd, \dots upper diagonal. Using this notation let us illustrate how the structure of the matrices look like:
for  $S_z \ge 0$
\[\{\hat M^{-1/2},\hat M^{1/2}\}=
\begin{cases}
\{\textsf{uud},\textsf{d}\}, & i \le 2S_z \\
\{\textsf{ud},\textsf{ld}\}, & i > 2S_z
\end{cases}
\]
and for $S_z < 0$
\[\{\hat M^{-1/2},\hat M^{1/2}\}=
\begin{cases}
\{\textsf{d},\textsf{lld}\}, & i \le 2|S_z| \\
\{\textsf{ud},\textsf{ld}\}, & i > 2|S_z|.
\end{cases}
\]

For completeness we also provide results for spin-1. For spin-1 systems $|S_z| \le N$, and we split $S_z$ into portions of 1 here. The conservation laws are:
\begin{eqnarray}
m_{i-1}&=&m_i+s_i-{\rm Sgn}(S_z) \cdot 1, ~~{\rm for} ~~ i \le |S_z| \\
m_{i-1}&=&m_i+s_i ~~~~{\rm for}~~~ i > |S_z|,
\end{eqnarray}
and the structure of the matrices is
for $S_z \ge 0$
\[\{\hat M^{-1},\hat M^{0},\hat M^{1}\}=
\begin{cases}
\{\textsf{uud},\textsf{ud},\textsf{d}\}, & i \le S_z \\
\{\textsf{ud},\textsf{d},\textsf{ld}\}, & i > S_z
\end{cases}
\]
and for $S_z < 0$
\[\{\hat M^{-1},\hat M^{0},\hat M^{1}\}=
\begin{cases}
\{\textsf{d},\textsf{ld},\textsf{lld}\}, & i \le |S_z| \\
\{\textsf{ud},\textsf{d},\textsf{ld}\}, & i > |S_z|.
\end{cases}
\]

In order to explicitly build up a U(1) symmetric matrix, one has to choose the dimensions of the degeneracy spaces for the bond dimensions. In practice, we have to take
a suitable set $D=\{t_{m_1},t_{m_2},\ldots,t_{m_n}\}$, where the $t_{m_i}$ denote the dimension of each degeneracy space. For a U(1) symmetric product state the choice would be $D=\{1\}$, and for an entangled state it may be, e.g., $D=\{1,1,3,3,1,1\}$.
The set $D$ is not determined by the symmetry and in principle many possibilities exist. There is no {\it a priori} principle which dictates a suitable choice.

In practical implementations one just has to ensure the specific block structure
of the matrices in order to maintain U(1) symmetry and obtain an MPS
with the desired spin projection $S_z$.
There are many ways to do this in practice, and details depend on the software system used
to implement the algorithm. In particular, the obvious sparseness of the matrices
must be employed in order to save computational resources. In our implementation
we use the sparse matrix technology available in {\it Mathematica} 10. This requires
very little programming effort. We only have to realize two facts: 1) the matrices $Q^{L,R}$ defined in Eq.~(\ref{LRnormalization}) are block diagonal for U(1) symmetric MPS, so the regauging
can be done blockwise; 2) the
generalized eigenvalue problem which must be solved in order to update a local matrix should contain only reduced matrix elements, i.e. rows and columns of zeros in $H_{\rm eff}$ and $N_{\rm eff}$
corresponding to the zeros of the MPS (caused by the block diagonal structure) must be removed before one starts to solve the eigenvalue problem.
After each update step we reconstruct the full structure of each local tensor
as a sparse tensor. Of course, one could implement the algorithm in terms of
reduced tensors only. But the sparse tensor technology employed here
saves resources in a similar way and is easier to implement.

\section{Application to the spin-1/2 XXZ model}~\label{XXZmodel}

Let us finally apply the algorithm developed above to a physically interesting model, the spin-1/2 XXZ model.
Due to the Mermin-Wagner theorem~\cite{PhysRevLett.17.1133} the continuous U(1) symmetry of the model cannot be broken, while the $Z_2$ symmetry is broken in the ferromagnetic and anti-ferromagnetic phases. As a consequence the magnetizations in the $x$ and $y$ axes direction vanish ($m_x=m_y=0$) as do the corresponding staggered magnetizations ($\bar{m}_x=\bar{m}_y=0$). Furthermore, for the correlators it holds, that $\langle s_i^x \otimes s_{i+1}^y\rangle=\langle s_i^y \otimes s_{i+1}^x\rangle=0$, and  $\langle s_i^x \otimes s_{i+1}^x\rangle=\langle s_i^y \otimes s_{i+1}^y\rangle$. We confirmed that all these relations hold numerically in our calculations.

Furthermore, due to the U(1) symmetry the $z$-magnetization $m_z=\langle S_z\rangle/N$
is a conserved quantity, which may be used to label the various states of the system. In fact, the  ground states in the XY and antiferromagnetic phases have $m_z=0$, while in the ferromagnetic phase the ground state has $m_z=1/2$. However, in the present paper we will study not only the ground state in the different phases, but also states with different $m_z$, e.g. the state with $m_z=0$ in the ferromagnetic region, which has interesting entanglement properties. But we will only study the ground state in different $m_z$ sectors.

The 2-spin reduced density matrix of the XXZ model may be easily expressed in terms of the spin correlators~\cite{PhysRevA.68.060301}, and in view of the U(1) symmetry the density matrix takes the following form
\begin{widetext}
\begin{equation}\label{rho12other}
 {\rho_{12}=\begin{pmatrix} \frac{1}{4}+\mathcal{Z}+m_z & 0 & 0 & 0 \\ 0 & \frac{1}{4}-\mathcal{Z}+\bar{m}_z & E-\Delta\mathcal{Z}-hm_z & 0 \\ 0 & E-\Delta\mathcal{Z}-hm_z & \frac{1}{4}-\mathcal{Z}-\bar{m}_z & 0 \\ 0 & 0 & 0 & \frac{1}{4}+\mathcal{Z}-m_z \end{pmatrix}}.
\end{equation}
\end{widetext}
To bring the density matrix into this form we used that $E=2\langle s_i^x \otimes s_{i+1}^x\rangle+\Delta \, \mathcal{Z}+hm_z$  due to Eq.~(\ref{XXZmodeleq}).
The density matrix $\rho_{12}$ is completely specified in terms of $E$, the magnetization $m_z$, the correlator  $\mathcal{Z}=\langle s_i^z \otimes s_{i+1}^z\rangle$, and the staggered magnetization $\bar{m}_z=\frac{1}{2} \langle s_i^z - s_{i+1}^z\rangle$. Numerically these quantities can be calculated using Eq.~(\ref{eq:matrixe}) and appropriate MPOs for each observable.

The single spin reduced density matrix is obtained as a partial trace of $\rho_{12}$ over the second site,
\begin{equation}\label{rho1}
\rho_1=\begin{pmatrix} \frac{1}{2}+m_z+\bar{m}_z & 0 \\ 0 & \frac{1}{2}-m_z-\bar{m}_z \end{pmatrix}.
\end{equation}
From this density matrix one immediately obtains the one-tangle,
\begin{equation}\label{tau1def}
\tau_1=4\det \rho_1= 1-4(m_z+\bar{m}_z)^2,
\end{equation}
which we will use as an entanglement quantifier of XXZ states. It characterizes the entanglement between one site and the rest of the system.

Other entanglement quantifiers we shall use are the concurrence of formation~\cite{PhysRevLett.80.2245} and the concurrence of assistance~\cite{Laustsen2003} defined as
\begin{eqnarray}\label{conc}
C_F(\rho_{12})&=&\max(0,\lambda_1-\lambda_2-\lambda_3-\lambda_4), \\
C_A(\rho_{12})&=&\lambda_1+\lambda_2+\lambda_3+\lambda_4,
\end{eqnarray}
where $\lambda_1,\lambda_2,\lambda_3,\lambda_4$ are the square roots of the eigenvalues (in decreasing order) of the non-Hermitian matrix $\rho_{12}\tilde{\rho}_{12}$ with $\tilde{\rho}_{12}=(\sigma_y\otimes\sigma_y)\rho_{12}^*(\sigma_y\otimes\sigma_y)$ and $\rho^*_{12}$ the complex conjugate of $\rho_{12}$. The concurrence of formation quantifies the nearest-neighbor two-site entanglement, while the concurrence of assistance measures the maximal bipartite entanglement which can be obtained while doing measurements on the rest of the spins.

In the following we will study properties of the spin-1/2 XXZ model as a function of the anisotropy parameter $\Delta$  for various $m_z$ using the formulas given above. Numerical calculations are presented for system sizes of $N=50$ and $N=100$ sites. The results are compared to analytical results as well as Bethe Ansatz calculations.

We test if the MPS determined by our algorithm is an eigenstate of the Hamiltonian by calculating   the variance per site $\delta H=\sqrt{\langle H^2 \rangle -\langle H \rangle ^2}/N$.  For an eigenstate it holds that $\delta H=0$.
The calculation of $\langle H^2 \rangle$ is briefly discussed in Appendix B.

\subsection{Correlators and entanglement properties at $h=0$}\label{resh0}

In the tables~\ref{data50delta0} and~\ref{data50delta1} we present results for the XXZ model without magnetic field calculated for periodic systems with $N=50$ spins at different $m_z$.
For $\Delta=0$ the obtained ground state energies are compared to the analytical result~\cite{PhysRevA.80.032102}
\begin{equation}\label{emz-fs}
E_0(m_z)=-\frac{1}{N \sin \frac{\pi}{N}} \, \cos(\pi m_z),
\end{equation}
while for $\Delta=1$ they are compared to finite system Bethe Ansatz calculations~\cite{KAR98}.
{The one-tangle and the concurrence of assistance indicate that entanglement monotonously grows from the product state $m_z=1/2$ to the state $m_z=0$, but nearest-neighbor entanglement peaks somewhere off $m_z=0$ for $\Delta=1$, thus indicating somewhat complicated entanglement structure.} For the entangled states $m_z<1/2$ a rather intricate choice of degeneracy set $D$ is required for a reasonable precision of the energy. This issue will be discussed in more detail in the following subsection.

Furthermore, in the tables we list our choice for the numbers $p$ and $p^{\prime}$ of singular values to be kept in the SVD of $N_L^{[i]}$, $N_R^{[i]}$, $H_L^{[i]}$, $H_R^{[i]}$. We ensure that the ratio (largest singular value)/(lowest kept singular value) is about $10^{-11}$ as recommended in~\cite{PIP10}. It holds that $p_{\rm max}=m^2$ and $p_{\rm max}^{\prime}=2m^2$ for the XXZ Hamiltonian (this can be obtained by Gauss elimination of the transfer matrix~\cite{weyrauchrakov}). We checked that for the spin-1 Heisenberg model the singular values decay very fast for systems of size $N=100$ as observed in Ref.~\cite{PIP10}. On the contrary, for spin-1/2 XXZ model a large percentage of singular values (at least 30\%) must be kept for a system of 100 sites. For sizes $N>150$ the parameters $p$, $p^{\prime}$ can be reduced roughly proportionally to ${1}/{N^2}$.
So $p$ and $p^\prime$ must be controlled carefully throughout the algorithm by monitoring the ratio (largest singular value)/(lowest kept singular value).

\begin{table*}[t]
\begin{tabular}{|c|c|c|c|c|c|c|c|c|c|c|}
  \hline
  $m_z$ & $E$ & $E_T$ & $\Delta E$ & $\mathcal{Z}$ & $\tau_1$ & $C_F$ & $C_A$ & degeneracy set& $m$ & $p~p^\prime$ \\
  \hline
  0.5 & 0 & 0 &  & 0.25 & 0 & 0 & 0 & $\{1,1,1\}$ & 3& 9~18 \\
  \hline
  0.4 & -0.098423 & -0.098428 & $5.1 \cdot 10^{-5}$ & 0.150316  & 0.359993 & 0.165020 & 0.231194 & $\{1 \times 4,2,2,3,3,3,2,2, 1 \times 4\}$ & 25 &~625~1250 \\
  \hline
  0.3 & -0.187144 & -0.187221 & $4.1 \cdot 10^{-4}$ & 0.055058  & 0.640004 & 0.263648 & 0.500524 & $\{1 \times 8, 2, 3 \times 8, 2, 1 \times 8\}$ & 44 & 1830~3600\\
  \hline
  0.2 & -0.257447 & -0.257688 & $9.3 \cdot 10^{-4}$ & -0.025888 & 0.840037 & 0.312642 & 0.754026 &$\{1 \times 8, 2, 3 \times 8, 2, 1 \times 8\}$ & 44 & 1480~2950\\
  \hline
  0.1 & -0.302792 & -0.302930 & $4.6 \cdot 10^{-4}$ & -0.081479 & 0.960063 & 0.334296 & 0.934246 &$\{1 \times 8, 2, 3 \times 8, 2, 1 \times 8\}$ & 44 & 1640~3200\\
  \hline
  0   & -0.318517 & -0.318519 & $6.3 \cdot 10^{-6}$ & -0.101456 & 1.000000 & 0.339946 & 1.000000 &$\{1 \times 3,2,3,4,5,5,5,4,3,2,1 \times 3\}$ & 39 & 1240~2490 \\
  \hline
\end{tabular}
\caption{\footnotesize The energy $E$, the spin correlator $\mathcal{Z}$ and the entanglement quantifiers $\tau_1$, $C_F$ and $C_A$ as functions of $m_z$ for spin-1/2 XXZ model of $50$ sites at $\Delta=0$,
$E_T$ is calculated according to Eq.~(\ref{emz-fs}), $\Delta E=(E-E_T)/E_T$. The staggered magnetization $\bar{m}_z$ is zero to a high precision. The dispersion $\delta H$ for $m_z=0$ is $3.5 \cdot 10^{-4}$.
  \label{data50delta0}}
\end{table*}
\begin{table*}[t]
\begin{tabular}{|c|c|c|c|c|c|c|c|c|c|c|}
  \hline
  $m_z$ & $E$ & $E_T$ & $\Delta E$ & $\mathcal{Z}$ & $\tau_1$ & $C_F$ & $C_A$  & degeneracy set& $m$ & $p~p^\prime$\\
  \hline
  0.5 & 0.25 & 0.25 & 0 & 0.25 & 0 & 0 & 0 & $\{1,1,1\}$ &3 &9,18\\
  \hline
  0.4 & 0.051743  & 0.051741  & $3.5 \cdot 10^{-5}$ & 0.150092  & 0.359998 & 0.179531 & 0.216983 & $\{1 \times 5,2,3 \times 6,2,1 \times 5\}$ & 32& ~1024~2048\\
  \hline
  0.3 & -0.134075 & -0.134268 & $1.4 \cdot 10^{-3}$ & 0.051847  & 0.639977 & 0.305155 & 0.462994 & $\{1 \times 8, 2, 3 \times 8, 2, 1 \times 8\}$& 44 & 1640~3200\\
  \hline
  0.2 & -0.291461 & -0.292021 & $1.9 \cdot 10^{-3}$ & -0.039533 & 0.839907 & 0.372763 & 0.710159 & $\{1 \times 8, 2, 3 \times 9, 2, 1 \times 8\}$& 47 & 1680~3335\\
  \hline
  0.1 & -0.401968 & -0.402081 & $2.9 \cdot 10^{-4}$ & -0.113493 & 0.959960 & 0.391109 & 0.912827 & $\{1 \times 8, 2, 3 \times 9, 2, 1 \times 8\}$& 47 & 1870~3690\\
  \hline
  0   & -0.443474 & -0.443477 & $6.7 \cdot 10^{-6}$ & -0.147826 & 1.000000 & 0.386944 & 1.000000 & $\{1,3,5,7,7,7,5,3,1\}$ & 39 & 1190~2360\\
  \hline
\end{tabular}
\caption{\footnotesize The energy $E$, the spin correlator $\mathcal{Z}$ and the entanglement quantifiers $\tau_1$, $C_F$ and $C_A$ as functions of $m_z$ for spin-1/2 XXZ model of $50$ sites at $\Delta=1$,
$E_T$ is calculated from Bethe Ansatz~\cite{KAR98}, $\Delta E=(E-E_T)/E_T$. The staggered magnetization $\bar{m}_z$ is zero to a high precision. The variance $\delta H$ for $m_z=0$ is $4.1 \cdot 10^{-4}$.
\label{data50delta1}}
\end{table*}

Analogous results for $N=100$ spins for $\Delta=0$ and $\Delta=1$ are presented in Tables~\ref{data100delta0} and \ref{data100delta1}, respectively. This system is already large enough that we can also compare
to infinite system Bethe Ansatz energies. For convenience, we briefly review the necessary formulas in the Appendix A. Infinite system Bethe Ansatz results are available analytically
and the whole phase diagram sketched in Fig.~\ref{phased} is easily obtained. Finite size Bethe Ansatz results are not available to us for the whole range of $\Delta$.
\begin{table*}[t]
\begin{tabular}{|c|c|c|c|c|c|c|c|c|c|c|c|}
  \hline
  $m_z$ & $E$ & $E_T$ & $E_{\infty}$ & $\Delta E$ & $\mathcal{Z}$ & $\tau_1$ & $C_F$ & $C_A$ & degeneracy set & $m$ & $p~p^\prime$\\
  \hline
  0.5 & 0 & 0 & 0 &  & 0.25 & 0 & 0 & 0 &  $\{1,1,1\}$ &3 & 9~18\\
  \hline
  0.4 & -0.097939 & -0.098379 & -0.098363 & $4.5 \cdot 10^{-3}$ & 0.150  & 0.360 & 0.158 & 0.237 & $\{1 \times 4,2,2,3\times 3,2,2,1 \times 4 \}$ & 25 &~290~~580\\
  \hline
  0.3 & -0.183044 & -0.187129 & -0.187098 & $2.2 \cdot 10^{-2}$ & 0.058  & 0.640 & 0.229 & 0.522 & $\{1 \times 8, 2, 3 \times 8, 2, 1 \times 8\}$& 44 &~590~1150\\
  \hline
  0.2 & -0.245605 & -0.257560 & -0.257518 & $4.6 \cdot 10^{-2}$ & -0.016 & 0.839 & 0.248 & 0.775 & $\{1 \times 8, 2, 3 \times 8, 2, 1 \times 8\}$ & 44&~970~1720\\
  \hline
  0.1 & -0.296970 & -0.302780 & -0.302731 & $1.9 \cdot 10^{-2}$ & -0.076 & 0.960 & 0.310 & 0.937 &$\{1 \times 8, 2, 3 \times 8, 2, 1 \times 8\}$&44 & 1120~2050\\
  \hline
  0   & -0.318340 & -0.318362 & -0.318310 & $6.9 \cdot 10^{-5}$ & -0.101 & 1.0 & 0.339 & 1.0 & $\{1,1,1,2,3,4,5\times 3,4,3,2,1,1,1\}$ & 39  &~835~1650\\
  \hline
\end{tabular}
\caption{\footnotesize The energy $E$, the spin correlator $\mathcal{Z}$ and the entanglement quantifiers $\tau_1$, $C_F$ and $C_A$ as functions of $m_z$ for spin-1/2 XXZ model of $100$ sites at $\Delta=0$,
$E_T$ is calculated according to Eq.~(\ref{emz-fs}), $\Delta E=(E-E_T)/E_T$. The staggered magnetization $\bar{m}_z$ is zero to a high precision. The variance $\delta H$ for $m_z=0$ is $5.2 \cdot 10^{-4}$.
  \label{data100delta0}}
\end{table*}

\begin{table*}[t]
\begin{tabular}{|c|c|c|c|c|c|c|c|c|c|c|c|}
  \hline
  $m_z$ & $E$ & $E_T$ & $E_{\infty}$ & $\Delta E$ & $\mathcal{Z}$ & $\tau_1$ & $C_F$ & $C_A$ & degeneracy set & $m$ & $p~ p^\prime$\\
  \hline
  0.5 & 0.25 & 0.25 & 0.25 & 0 & 0.25 & 0 & 0 & 0 & $\{1,1,1\}$ & 3 & ~9~18\\
  \hline
  0   & -0.443205 & -0.443230 & -0.443147 & $5.7 \cdot 10^{-5}$ & -0.148 & 1.0 & 0.386 & 1.0 & $\{1,3,5,7\times 3,5,3,1\}$ & 39 & 725~1450\\
  \hline
\end{tabular}
\caption{\footnotesize The energy $E$, the spin correlator $\mathcal{Z}$ and the entanglement quantifiers $\tau_1$, $C_F$ and $C_A$ as functions of $m_z$ for spin-1/2 XXZ model of $100$ sites at $\Delta=1$,
$E_T$ is calculated from Bethe Ansatz~\cite{KAR98}, $\Delta E=(E-E_T)/E_T$. The staggered magnetization $\bar{m}_z$ is zero to a high precision. The variance $\delta H$ for $m_z=0$ is $6.6 \cdot 10^{-4}$.
  \label{data100delta1}}
\end{table*}

\begin{figure}[h!]
    \includegraphics[width=0.45\textwidth]{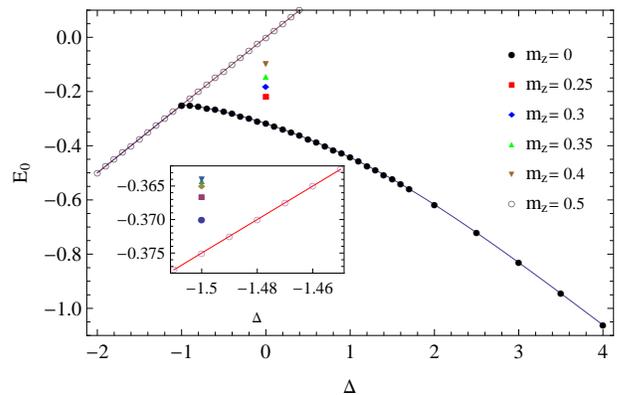}
    \caption{\footnotesize (color online)
    Energy per site $E(m_z)$ for a spin-1/2 XXZ ring of 100 sites at zero magnetic field as a function of the anisotropy parameter $\Delta$. The full line shows the Bethe Ansatz result Eq.~(\ref{e0Bethe}). The various symbols described in the legend correspond to numerical results for selected $m_z$. The inset shows results at $\Delta=-1.5$ for $m_z=.5$, .49, .48, .47, .46, .45.} \label{energyh0}
\end{figure}

In Fig.~\ref{energyh0} we compare the infinite size Bethe Ansatz energies with numerical results for $N=100$ in the parameter interval $-2<\Delta<4$. The ground state energy at $m_z=0$ for $\Delta\ge -1$ agrees with Bethe Ansatz results up to finite-size corrections  $\Delta E/E \sim 10^{-4}$.
For $\Delta\le-1$ and infinite system size the ground state energies $E_0(m_z)={\Delta}/{4}$ are independent of
$m_z$~\cite{PhysRev.147.303}, which means that in this parameter region the ground state is infinitely degenerate.
The degeneracy of the states with different $m_z$ is obtained numerically at $\Delta=-1$ with high precision. 
%
However, for finite systems the degeneracy is lifted  for $\Delta<-1$  (Fig.~\ref{energyh0} inset). The energy per site of the state with $m_z=1/2-1/N$ is given by an exact solution
\begin{equation}~\label{mzlarge}
E_0=\Delta/4+(|\Delta|-1)/N
\end{equation}
indicating a quite significant finite site effect at rather moderate $\Delta$.
The corresponding numerical result shown in Fig.~\ref{energyh0} (inset)  exactly agrees with Eq.~(\ref{mzlarge}).
In addition, in the inset of Fig.~\ref{energyh0} we show results for a few other states with large magnetization
which show even larger finite size effects.

\begin{figure}[h!]
    \includegraphics[width=0.45\textwidth]{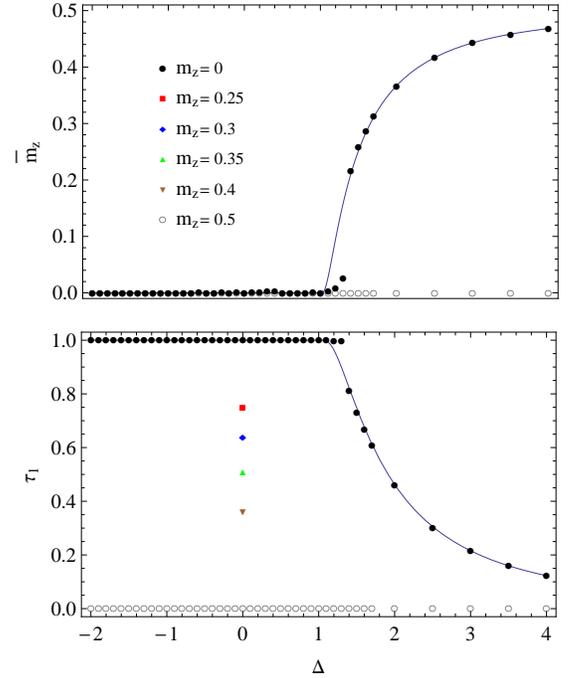}
    \caption{\footnotesize (color online)
    Staggered magnetization per site $\bar{m}_z$ (top) and one-tangle $\tau_1$ (bottom) of the 1D spin-1/2 XXZ ring of 100 sites at zero magnetic field as a function of the anisotropy parameter $\Delta$. Numerical results (symbols) are compared to results calculated from Eqs.~(\ref{Bethe-mzst}) and (\ref{tau1def}). Significant finite-size effects are observed for $1\le \Delta \lesssim 1.4$.
    } \label{magzstagtau1h0}
\end{figure}

In Fig.~\ref{magzstagtau1h0} the staggered magnetization $\bar{m}_z$ is displayed for a system with $N=100$ compared to the Bethe Ansatz result given in the Appendix A. As expected, one finds that the staggered magnetization is non-zero only in the anti-ferromagnetic region $\Delta>1$. For $1\le \Delta \lesssim 1.4$ we observe large finite-size effects.
In addition, we show in Fig.~\ref{magzstagtau1h0} results for the one-tangle calculated from $\bar{m}_z$
using Eq.~(\ref{tau1def}). The result indicates that the $m_z=0$ state is strongly entangled for $\Delta<1$. Above
$\Delta=1$ this state slowly `looses' entanglement with increasing $\Delta$. In order to obtain correct numerical results for the staggered magnetization at $-1<\Delta<1$ it is important that U(1) symmetry is preserved. Typically non-symmetric codes obtain spurious results for $\bar{m}_x$ (which becomes nonzero) and consequently for $\tau_1$.

\begin{figure}[h!]
\includegraphics[width=0.45\textwidth]{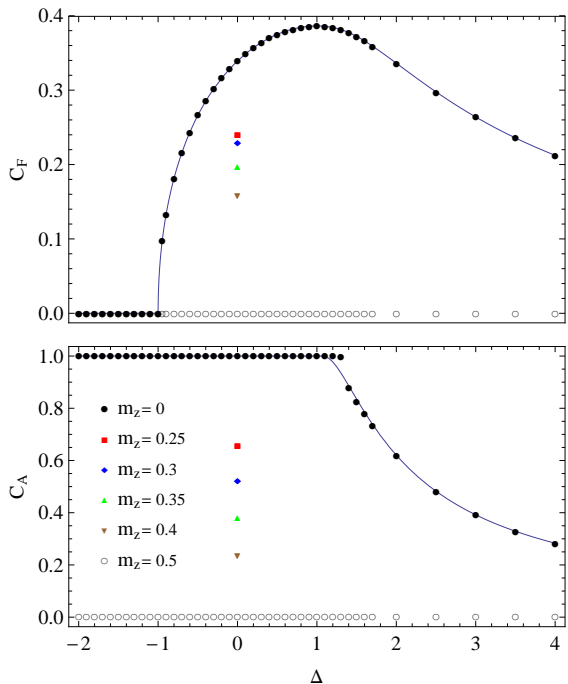}
    \caption{\footnotesize (color online)
    (top) Concurrence of formation $C_F$ of 1D spin-1/2 XXZ ring of 100 sites at zero magnetic field as a function of the anisotropy parameter $\Delta$ compared to Bethe Ansatz results (full line).
    (bottom) Concurrence of assistance $C_A$ of 1D spin-1/2 XXZ ring of 100 sites at zero magnetic field as a function of the anisotropy parameter $\Delta$ compared to Bethe Ansatz results (full line). Significant finite-size effects are observed for $1\le \Delta \lesssim 1.4$.
    } \label{concurrence-h0}
\end{figure}

Finally, we present results for the concurrence of formation $C_F$ and the concurrence of assistance $C_A$ for a system of 100 sites in Fig.~\ref{concurrence-h0} again compared to infinite size Bethe Ansatz results. One observes for $C_A$ large finite size effects close to the critical point at $\Delta=1$. The trace for the concurrence of assistance $C_A$ looks very similar to that of $\tau_1$. However, the concurrence of formation $C_F$ shows a very different characteristic as it is maximal at $\Delta=1$
and zero for $\Delta<-1$. In this respect the $m_z=0$ XXZ state for $\Delta<-1$ is similar to the Greenberger-Horne-Zeilinger (GHZ) state, which has zero concurrence of formation but is highly entangled with
one-tangle or concurrence of assistance equal to 1.

There is a somewhat indirect quantification of entanglement: the bond size $m$ of the matrices of the MPS as given
in the tables.
The required bond sizes $m$ for states with large but not full magnetization indicate that these states are characterized by entanglement not measured by the simple quantifiers $\tau_1$, $C_F$, or $C_A$. { Long-ranged entanglement or many-way entanglement may be a better way to quantify the entanglement of these states.}

\subsection{Accuracy and precision of the algorithm}\label{precision}

The results presented in the previous section are meant to illustrate the algorithm, and we
did not attempt to push the calculations to the limit in order to obtain the best possible accuracy. Nevertheless, with relatively small MPS sizes one obtains results in quite good agreement with other approaches.

As is obvious from the results, the accuracy depends crucially  on the chosen degeneracy set, which also determines the overall MPS size $m$.  Of course, since the algorithm is variational, it entails an iterative minimization, and the number of iteration steps taken is another important parameter. Often we can easily increase the precision of our results by
adopting more stringent convergence requirements at the expense of a longer computing time.
For the present paper we stopped our numerical update (i.e. minimization) procedure if the averaged relative ground state energy does not change more than
$10^{-7}$ within the last $N/3$ update steps of the algorithm.
However, it is possible that for a given degeneracy set $D$ the approach to the minimum may be excessively slow, and the optimization stops before reaching the minimum. Moreover, occasionally the algorithm may get stuck in a local minimum.

It would be desirable that the algorithm chooses an optimal degeneracy set $D$ automatically.
For OBC such a procedure exists, and we will briefly review this method here. It was introduced
by White~\cite{WHI05} and entails a modification of the regauging step.
Instead of Eqs.~(\ref{LRnormalization}) the following constructions are calculated,
\begin{eqnarray}\label{LRdensitymatrix}
\rho_{(a_{i-1},s_i),(a_{i-1}^{\prime},s_i^{\prime})}=\sum_{a_i} M^{[i],s_i}_{a_{i-1},a_i} M^{[i],s_i^{\prime} *}_{a_{i-1}^{\prime},a_i}~~ \text{left-n.}, \nonumber\\
\rho_{(s_i,a_i),(s_i^{\prime},a_i^{\prime})}=\sum_{a_{i-1}} M^{[i],s_i}_{a_{i-1},a_i} M^{[i],s_i^{\prime} *}_{a_{i-1},a_i^{\prime}}~~ \text{right-n.},~~~
\end{eqnarray}
$\rho$ has  size $(2s+1)m \times (2s+1)m $. Its SVD  $\rho=USV^{\dag}$ has exactly $m$ singular values, and therefore $\rho$ has matrix rank $m$, and one obtains the regauged matrix $M^{[i],s_i}_{a_{i-1},a_i}=U_{(a_{i-1},s_i),a_i}$. This matrix is identical to the one obtained from Eqs.~(\ref{LRnormalization}). It can be shown that for OBC the constructions~(\ref{LRdensitymatrix}) corresponds to the reduced  density matrix $\rho_{1 \rightarrow i}$ for the sites from 1 to $i$ (left-normalization) and $\rho_{i \rightarrow N}$ for the sites from $i$ to $N$ (right-normalization), respectively. They can be calculated here from a single tensor $M^{[i]}$.

In the case of U(1) symmetry $\rho$ is block diagonal with each block corresponding to a quantum number $m_{i-1}-s_i$ (left-normalization) or $s_i+m_i$ (right-normalization). Due to the `conservation laws' each nonzero block of $\rho$ corresponds to an analogous block of $Q^{L,R}$, and corresponding blocks have the same rank. Thus regauging can be done block-wise, and the same results are obtained as if Eqs.~(\ref{LRnormalization}) were used.

The crucial step proposed by White~\cite{WHI05} for OBC is a modification of the density matrix $\rho$ (see, e.g.~Eq.~(217) in Ref.~\cite{SCH10}). The matrix rank of the modified density matrix is larger than $m$. Again one calculates an SVD of this matrix $\rho=USV^{\dag}$ and constructs the regauged local tensor $M^{[i]}$ from the matrix $\tilde{U}$ corresponding to the $m$ largest singular values. For U(1) symmetric MPS one selects the $m$ largest singular values irrespective to which degeneracy sector they belong. In this way degeneracy sectors may increase or decrease in size or sectors may even be lost or created dynamically during the optimization procedure.

Unfortunately, this procedure does not work for PBC:
The reduced density matrix $(\rho_{1 \rightarrow i})$ (needed for left-normalization) is for both OBC and PBC
given by
\begin{multline}\label{densitymatrixPBC}
(\rho_{1 \rightarrow i})_{(s_1,\cdots,s_i),(s_1^{\prime},\cdots,s_i^{\prime})}=\\
{\rm Tr}((M^{[1],s_1} \otimes M^{[1],s_1^{\prime *}})\cdots (M^{[i],s_i} \otimes M^{[i],s_i^{\prime *}}) \cdot N_R^*),
\end{multline}
and in general it involves all MPS tensors. However, for OBC  $N_R=1$ and, as alluded to above, the rank of $\rho$ is only $m$, and $\rho$ can be written in terms of a single tensor $M^{[i]}$.  This simplification does not happen for PBC, and the reduced density matrix has size and rank $d^i$.

From these considerations we see that the construction of an algorithm for the selection of degeneracy sets for PBC faces different issues than for OBC, and we here opted to determine them by numerical tests as was
also done by Vidal and collaborators~\cite{ SIN2012} for U(1) symmetric MERA implementations.
As a consequence, an alternative to the approach proposed in~\cite{WHI05} for OBC is desirable, but beyond the scope of the present paper.

\section{Conclusion}

In this paper we propose a specific new way to construct U(1) covariant MPS for PBC and discuss many aspects concerning the construction of symmetric MPS not covered elsewhere. We implement our proposal in a variational algorithm for finite spin systems based on the PBC algorithm of Verstaete, Porras, and Cirac~\cite{VER04} as modified by Pippan, White, and Evertz~\cite{PIP10}.

The algorithm is applied to a study of the properties of the  spin-1/2 XXZ model for systems of 50 and 100 sites. It proves to be numerically stable, and our results agree rather well with predictions of the Bethe Ansatz. The algorithm correctly captures the properties of the system in the XY phase, where other numerical algorithms break the U(1) symmetry.

The convergence properties of the proposed algorithm are studied. Our concrete choice of appropriate U(1) degeneracy sectors is provided, and we exemplify that the replacement of long products of transfer matrices by their truncated singular value decomposition (SVD) must be used with caution. We demonstrate, that one must keep many more singular values than for spin-1 systems discussed in Ref.~\cite{PIP10}.

We calculate various spin correlation functions and entanglement quantifiers for the XXZ model as a function of the anisotropy parameter $\Delta$ and the magnetization $m_z$ at zero magnetic field. We show analytically and numerically that entanglement in general decreases monotonically with increasing magnetization of the system. The concurrence of formation shows a deviation from this rule for systems with small magnetization.

The present work could be extended in many ways. Most importantly a general algorithmic strategy to choose the appropriate degeneracy sectors is needed. In this way it may be also possible to improve the numerical results for intermediate spin projections $m_z$. Such work is presently under way as a generalization of the proposal made by White~\cite{WHI05} for OBC.

\appendix

\section*{Appendix A: Infinite size Bethe Ansatz results}\label{append}

The energy per site of the $m_z=0$ state as determined by the infinite size Bethe Ansatz~\cite{PhysRev.147.303,PhysRev.150.327} is given by

\begin{eqnarray}
E_0&=&\frac{\Delta}{4} ~~~~~~~~~~~~~~~~~~~~~~~~~~~~~~~~ {\rm for }~~  \Delta\leq-1, \\
E_0&=&\frac{\Delta}{4}-\frac{1}{2}(1-\Delta^2)\times ~~~~~~~~~~~{\rm for }~~\Delta > -1 \label{e0Bethe}\\
   & & ~~\int\limits_{-\infty}^{\infty}\frac{dx}{\cosh \pi x (\cosh (2x\arccos\Delta)-\Delta)}. \nonumber
 \end{eqnarray}
At $\Delta=1$ the integrand is not well defined, and one needs to take an appropriate limit.
One obtains the $\mathcal{Z}$ correlator from a derivative of the energy with respect to $\Delta$.

The staggered magnetization $\bar{m}_z$ is given by~\cite{baxter73},
\begin{eqnarray}\label{Bethe-mzst}
\bar{m}_z=0 ~~~{\rm if}~ \Delta<1,\nonumber\\
\bar{m}_z=\frac{1}{2}\prod_{n=1}^\infty \tanh^2(n~ {\rm arccosh}\,\Delta)~~~~{\rm if}~ \Delta\geq 1.
\end{eqnarray}
From these results the complete density matrix Eq.~(\ref{rho12other}) can be determined, which enables the
calculation of the entanglement quantifiers discussed in section~\ref{XXZmodel}.

\section*{Appendix B: Calculation of $\langle H^2 \rangle$ }\label{append2}

The MPO for the calculation of $\langle H^2 \rangle$ is given by
$$\mathcal{W}_{(b_{i-1} b_{i-1}^{\prime}),(b_i b_i^{\prime})}^{[i],s_i,s_i^{\prime}}=\sum_{s_i^{\prime\prime}} W_{b_{i-1},b_i}^{[i],s_i,s_i^{\prime\prime}} \, W_{b_{i-1}^{\prime},b_i^{\prime}}^{[i],s_i^{\prime\prime},s_i^{\prime}}.$$
This MPO represents a matrix of size $25 \times 25$ for the XXZ model (its explicit form is not written down due to its large size); $\langle H^2 \rangle$ can be obtained from this MPO using Eq.~(\ref{eq:matrixe}) and appropriate transfer matrices~$E_{\mathcal{W}}^{[i]}$ of size $25m^2 \times 25m^2$.

However, it can be shown explicitly by Gauss elimination that all the transfer matrices (and consequently their products) have only $p^{\prime\prime}=6m^2$ nonzero singular values. So the multiplication of $N$ transfer matrices can be done by the efficient update proposed in~\cite{PIP10,weyrauchrakov}.

\begin{acknowledgments}
We thank Ian P. McCulloch for a useful correspondence. Mykhailo V. Rakov thanks Physikalisch-Technische Bundesanstalt for financial support during short visits to Braunschweig.
\end{acknowledgments}

\end{document}